\documentclass[journal]{IEEEtran}

\usepackage{cite}
\usepackage{url}

\ifCLASSINFOpdf
\else
\fi

\usepackage{graphicx} 
\usepackage{comment}

\usepackage{amsmath, amssymb}

\ifCLASSOPTIONcompsoc
\usepackage[caption=false,font=normalsize,labelfont=sf,textfont=sf]{subfig}
\else
\usepackage[caption=false,font=footnotesize]{subfig}
\fi

\usepackage[dvipsnames]{xcolor}

\usepackage{epstopdf}

\begin{document}
	\title{Energy efficiency of slotted LoRaWAN communication with out-of-band synchronization}
	
	
\author{\vspace{-25pt}}
\author{Luca~Beltramelli,~\IEEEmembership{Member,~IEEE,}
	Aamir~Mahmood,~\IEEEmembership{Senior~Member,~IEEE,}
	Patrik~{\"O}sterberg,~\IEEEmembership{Member,~IEEE,}
	Mikael~Gidlund,~\IEEEmembership{Senior~Member,~IEEE,}
	Paolo~Ferrari,~\IEEEmembership{Member,~IEEE,}
	and~Emiliano~Sisinni,~\IEEEmembership{Member,~IEEE}
	\thanks{L.~Beltramelli, A.~Mahmood, P.~{\"O}sterberg, and M.~Gidlund  are with the Department of Information Systems and Technology, Mid Sweden University, 851~70~Sundsvall, Sweden, e-mail: luca.beltramelli@miun.se.
	P.~Ferrari and E.~Sisinni are with the Department of Information Engineering, University of Brescia, 25123 Brescia, Italy, email:emiliano.sisinni@unibs.it.}
	\vspace{-25pt}}

	\maketitle
	
	\begin{abstract}
		Although the idea of using wireless links for covering large areas is not new, the advent of Low Power Wide Area Networks (LPWANs) has recently started changing the game. Simple, robust, narrowband modulation schemes permit the implementation of low-cost radio devices offering high receiver sensitivity, thus improving the overall link budget. The several technologies belonging to the LPWAN family, including the well-known LoRaWAN solution, provide a cost-effective answer to many Internet-of-things (IoT) applications, requiring wireless communication capable of supporting large networks of many devices (e.g., smart metering). Generally, the adopted medium access control (MAC) strategy is based on pure ALOHA, which, among other things, allows to minimize the traffic overhead under constrained duty cycle limitations of the unlicensed bands. Unfortunately, ALOHA suffers from poor scalability, rapidly collapsing in dense networks. This work investigates the design of an improved LoRaWAN MAC scheme based on slotted ALOHA. In particular, the required time dissemination is provided by out-of-band communications leveraging on Radio Data System \mbox{(FM-RDS)} broadcasting, which natively covers wide areas both indoor and outdoor. An experimental setup based on low-cost hardware is used to characterize the obtainable synchronization performance and derive a timing error model. Consequently, improvements in success probability and energy efficiency have been validated by means of simulations in very large networks with up to 10000 nodes. It is shown that the advantage of the proposed scheme over conventional LoRaWAN communication is up to 100\% when short update time and large payload are required. Similar results are obtained regarding the energy efficiency improvement, that is close to 100\% for relatively short transmission intervals and long message duration; however, due to the additional overhead for listening the time dissemination messages, efficiency gain can be negative for very short duration of message fastly repeating.
	\end{abstract}
	
	
	\IEEEpeerreviewmaketitle

\section{Introduction}
	\label{Introduction}
	\IEEEPARstart{T}{he}  medium access control (MAC) mechanism is the most relevant part of the data link layer; placed just above the physical layer, it controls how the common network resources are shared. For this reason, it is one of the main control knobs of any communication solution, which must fully exploit the capabilities of the underlying physical layer without impairing application requirements.
	
	When typical (wireless) Internet-of-things (IoT) applications are considered~\cite{MACSurvey}, (i.e., for transmitting data from many simple and autonomous devices deployed in the field towards the Cloud), the emphasis is on simplicity and security, while other aspects are somewhat overshadowed. Distributed MAC approaches are generally preferred since they minimize the traffic overhead, avoiding the need for forwarding control information towards a single node. In particular, for non-critical IoT applications, random access paradigms are often used. 
	Although it is an inefficient mechanism, pure ALOHA is still widely used due to its many advantages; packets can have variable size, nodes can start transmission at any time, and time synchronization is not required. For all these reasons, pure ALOHA has been selected as the MAC protocol for most (if not all) Low Power Wide Area Networks (LPWANs), including the broadly adopted LoRaWAN~\cite{LPWANAloha}. However, the main limitation of ALOHA is the large number of collisions that can occur when the network consists of a large number of devices. Better performance could be obtained using the "listen-before-talk" (LBT) approach, as in carrier sense multiple access with collision avoidance (CSMA/CA). However, the effects of overhearing, hidden nodes, and long ongoing transmission detection can negate the theoretical performance improvement for low data rate communication over wide area.

	The transmission vulnerability time of ALOHA can be halved, assuming all the devices are relatively synchronized, by aligning the transmissions to slot boundaries as in slotted ALOHA (S-ALOHA)~\cite{SAlohaM2M}. 
	Such an approach could be particularly interesting for applications requiring periodic sampling of the devices, which must be somehow time synchronized~\cite{rizzi2017evaluation}, as common in most IoT-like monitoring applications such as smart metering~\cite{SmartMeter}.
	Thus, the time-dissemination strategy required by the application layer (which must schedule the sampling activity), can also be used by the lower layers of the communication protocol stack. 
	As a matter of fact, some researchers have already proposed the use of simple, in-band synchronization mechanisms for enhancing LoRaWAN communication~\cite{reynders2018improving, polonelli2019slotted, rizzi2019syncunc, haxhibeqiri2018low}. 
 
	However, the applications of interest mostly rely on uplink messages (from the field toward a data sink) and should aim at limiting the traffic overhead, especially in downlink (e.g., for exchanging time-related information), in order to minimize the risk of collisions and increase the scalability. As a consequence, the overhead imposed by in-band synchronization can result into intolerable reduction of the communication opportunities for the devices. On the contrary, out-of-band mechanisms could provide a more energy-efficient and effective (i.e., offering high capacity/device density) approach under the low duty-cycle restrictions imposed on LoRaWAN. 
	
	This work aims to propose and evaluate the performance of an S-ALOHA scheme for LoRaWAN using an out-of-band synchronization technology. In particular, the main contributions are:
	\begin{itemize}
		\item analysis of time-dissemination technologies able to cover wide area including indoor and outdoor scenarios with focus on FM-Radio Data System (\mbox{FM-RDS}),
		\item theoretical modeling and experimental evaluation of the timing errors in a real-world implementation,
		\item comparative analysis of simulation results between LoRaWAN and the proposed communication mechanism in terms of transmission success probability and energy efficiency for large-scale deployment in an urban scenario.
	\end{itemize}

	The rest of the paper is organized as follows. The details of LoRaWAN and time-dissemination technologies are presented in Sec. \ref{RelatedTechnologies}. The proposed S-ALOHA approach is described in Sec. \ref{sec:ProposedApproach}, while timing errors are analyzed in Sec. \ref{TimingErrors}. Experiments are detailed in Sec. \ref{Experimental Results}, while simulation results are reported in Sec. \ref{Simulator Results}. Finally, conclusions are drawn in Sec. \ref{sec:conclusions}.
	
	\section{LoRaWAN and the out-of-band synchronization}
	\label{RelatedTechnologies}
	As stated earlier, LPWANs are developed to provide multi-km network coverage for wireless IoT applications. They typically employ single-hop connectivity and are designed in order to minimize power consumption at the expense of data rate (of few kbps) and of latency (in the order of seconds).
    From the application point of view, the main target of LPWANs is to support delay-tolerant massive machine-type communications (mMTC). 
    Most mMTC applications have some common traffic characteristics; in particular, each device generates small and possibly infrequent messages, imposing a much higher traffic on uplink \cite{LPWANapps}. 
    On the contrary, with the exception for adaptive communication parameters update, downlink is seldomly used.
    Examples of such uplink-centric applications, implementing "fire-and-forget" communication strategy, include smart metering, fleet and asset tracking, vending machines management. Currently, the term LPWAN includes many different technologies, standardized or proprietary, operating in licensed or licence-free bands. The solutions that can boast the widest adoption and provide similar features are arguably NB-IoT, SigFox, and LoRa/LoRaWAN. Their most important characteristics are summarized in Table~\ref{tb:comparison}. For a comprehensive comparison of these technologies, we refer the reader to~\cite{comparison}.
	For the sake of completeness, note that 3GPP also supports LTE Cat-M, which is a solution fully compliant with LTE that leverages on reduced channel bandwidth to offer raw throughput in the order of few Mbps.

\begin{table}[!htb]
	\caption{Salient Features of different LPWAN Solutions}
	\centering
	\scalebox{0.97}{
	\renewcommand{\arraystretch}{0.7} 
	\begin{tabular}{|p{0.1\textwidth}|p{0.1\textwidth}|p{0.1\textwidth}|p{0.1\textwidth}|}
		\hline
		{\color[HTML]{000000} \textbf{Features}} & {\color[HTML]{000000} \textbf{Sigfox}} & {\color[HTML]{000000} \textbf{LoRaWAN}} & {\color[HTML]{000000} \textbf{NB-IoT (NB2)}} \\\hline
		{\color[HTML]{000000} \textbf{Modulation}} & {\color[HTML]{000000} BPSK} & {\color[HTML]{000000} CSS} & {\color[HTML]{000000} QPSK}\\
		{\color[HTML]{000000} \textbf{Spectrum}} & {\color[HTML]{000000} Unlicensed ISM bands} & {\color[HTML]{000000} Unlicensed ISM bands} & {\color[HTML]{000000} Licensed LTE bands}\\
		{\color[HTML]{000000} \textbf{Bandwidth}} & {\color[HTML]{000000} 100 Hz} & {\color[HTML]{000000} 250/125 kHz} & {\color[HTML]{000000} 180 kHz}\\
		{\color[HTML]{000000} \textbf{Max data rate}} & {\color[HTML]{000000} 100 bps} & {\color[HTML]{000000} 50 kbps} & {\color[HTML]{000000} 127 kbps (DL) / 159 kbps (UL)}\\
		{\color[HTML]{000000} \textbf{Bidirectional}} & {\color[HTML]{000000} Limited (Half-duplex)} & {\color[HTML]{000000} Yes (Half-duplex)} & {\color[HTML]{000000} Yes (Half-duplex)}\\
		{\color[HTML]{000000} \textbf{Max messages/day}} & {\color[HTML]{000000} 140 (UL), 4 (DL)} & {\color[HTML]{000000} Duty-cycle constrained} & {\color[HTML]{000000} Unlimited}\\
		{\color[HTML]{000000} \textbf{Max payload length}} & {\color[HTML]{000000} 12 bytes (UL), 8 bytes (DL)} & {\color[HTML]{000000} 242 bytes} & {\color[HTML]{000000} 2536 bit }\\
		{\color[HTML]{000000} \textbf{Interference immunity}} & {\color[HTML]{000000} Very high} & {\color[HTML]{000000} Very high} & {\color[HTML]{000000} Low}\\
		{\color[HTML]{000000} \textbf{Authentication \& encryption}} & {\color[HTML]{000000} Yes (Enc. is optional, AES128)} & {\color[HTML]{000000} Yes (AES128)} & {\color[HTML]{000000} Yes (LTE enc.)}\\
		{\color[HTML]{000000} \textbf{Adaptive data rate}} & {\color[HTML]{000000} No} & {\color[HTML]{000000} Yes} & {\color[HTML]{000000} No}\\
		{\color[HTML]{000000} \textbf{Handover}} & {\color[HTML]{000000} End-devices do not join BS} & {\color[HTML]{000000} End-devices do not join to BS} & {\color[HTML]{000000} End-devices join single BS}\\
		{\color[HTML]{000000} \textbf{Private nwk}} & {\color[HTML]{000000} No} & {\color[HTML]{000000} Yes} & {\color[HTML]{000000} No}\\
		{\color[HTML]{000000} \textbf{Standardization}} & {\color[HTML]{000000} Sigfox / ETSI} & {\color[HTML]{000000} LoRa-Alliance} & {\color[HTML]{000000} 3GPP}\\
		\hline
	\end{tabular}}
	\label{tb:comparison}
	\vspace{-10pt}
\end{table}
	
	In the following, a brief overview of LoRaWAN is provided, followed by a discussion on viable out-of-band time-dissemination technologies offering wide-area coverage.

	\subsection{LoRaWAN: A brief overview}
	\label{sec:LoRaWAN}
	LoRaWAN is a member of the LPWANs family, which includes all the communication solutions aimed at transferring limited amount of data over a wide area with relaxed time constraints. LPWANs generally operate in the sub-GHz unlicensed bands (that provides better propagation compared to the crowded $2.4$~GHz band) and exploit the reduced channel bandwidth, thus trading smaller data rates for increased receiver sensitivity. LoRaWAN is based on a proprietary radio technology (named LoRa), leveraging on an efficient chirp spread spectrum modulation. The poor bandwidth availability is overcome by quasi-orthogonal virtual channels, offering different data rates. These virtual channels depend on the spreading factor (SF): a tunable parameter, which modifies the symbol duration and represents the number of bits coded per symbol. In particular, given the SF, the chirp length is defined (affecting the data rate as well), and $2^{\textrm{SF}}$ different frequency trajectories exist. Additional interleaving and whitening strategies are applied to increase interference robustness, and forward error correction with coding rate $\textrm{CR}\in[4/5,..,4/8]$ is adopted as well. LoRa offers so many advantages that it is the foundation of many proprietary solutions (e.g., \cite{MultiChHop, LargeArea}), other than LoRaWAN, which provides a standardized protocol stack.
	
	The LoRaWAN channel access mechanism is pure ALOHA, chosen in order to minimize stack complexity and traffic overhead for communication management. However, it is well-known that ALOHA can be a significant bottleneck to the LoRaWAN scalability in dense scenarios. For this reason, in the literature some advanced random and scheduled access solutions have been discussed \cite{haxhibeqiri2018low, reynders2018improving, RT_LoRa, FREE}. The possible use of LBT, referred to as channel activity detection (CAD), is rarely adopted due to the limits mentioned earlier. Scheduling of transmissions reduces message collisions, however, it requires additional overhead for providing time synchronization across the whole network.
	
	From the topology point of view, LoRaWAN implements a star-of-stars architecture, in which GWs forward every incoming wireless message to the backhaul network and vice versa. The GWs do not manipulate user data, and the node identification and authentication are managed by the Network Servers (NS), whereas user data are managed by the Application Server (AS). LoRa devices offer different features, depending on their class; Class-A permits event-based uplink followed by an (optional) downlink, Class-B offers synchronized downlink, lastly, Class-C implements continuous listening. For the sake of clarity, in the rest of the paper, the term LoRaWAN refers to the complete communication solution, whereas LoRa to the physical layer.
	
	\subsection{Large areas out-of-band time-dissemination technologies}
	\label{sec:Time_Diss_Tech}
	A node executes a synchronization algorithm to align the time of its local clock to the clocks of other nodes or to a global clock.
	In particular, absolute synchronization requires the distributed clocks to refer to an actual global time standard (e.g., the Universal Coordinated Time UTC), whereas relative synchronization requires time alignment within the local system, involving only local nodes. In most practical cases, absolute synchronization is not a mandatory requirement.
	
	Synchronization implies transmission of time-related information and, when it occurs over wireless networks, must respect many constraints dictated by limited resource availability (energy, bandwidth, computation capability), network topology, and unreliable links. The cost factor is also important when the application scales. In the case of LPWANs, offering poor throughput and reduced number of physical channels, the constraints become more stringent. As a result, the overhead of two-way communication for in-band synchronization mechanisms can unacceptably reduce communication opportunities (e.g., due to limitations on the duty-cycle) \cite{haxhibeqiri2018low} especially for the aforementioned "fire-and-forget", uplink-centric, monitoring applications. For this reason, this work evaluates the out-of-band approach (sometimes referred to as hardware-assisted synchronization \cite{HWassisted}), where a complementary communication system is used (possibly wireless, as for wake-up receivers \cite{Wakeup}). The criteria for the selection of suitable synchronization technology for LPWANs are: a) accurate event timestamping for minimizing the local clock offset with respect to the chosen time reference (in other words, the complementary system must minimize the uncertainty of the time of arrival estimation,
	 b) wide-area coverage complemented by indoor/outdoor capability, and c) minimal energy consumption overhead. 
	
	In this work, we consider the following out-of-band technologies: global navigation satellite system (GNSS), radio controlled clocks (RCCs), and the Radio Data System \mbox{(FM-RDS)}.  
	\subsubsection{GNSS - GPS}
	The Global Positioning System (GPS) is the most diffused GNSS, which relies on strict time synchronization for evaluating pseudoranges in trilateration. Nowadays, GPS receivers can provide a 1-PPS time reference signal while offering excellent short-term stability and sub-microsecond synchronization accuracy \cite{GPSAccuracy}. Unfortunately, the cost of GPS receivers may be too high for several IoT applications, and their power consumption may not be compatible with battery-operated devices \cite{GPSPower}. Moreover, GPS receivers require line-of-sight with satellites and thus cannot be used indoor.  
	\subsubsection{RCC - DCF77}
	The long-wave RCCs operate in the $40$~kHz to $80$~kHz frequency bands. The use of low-frequency signaling offers wide-area coverage, lower power than typical radio frequency signals, and better indoor reception. 
	The DCF77 transmitter is operated by Physikalisch Technische Bundesanstalt in Frankfurt/Main, which covers all the central Europe. In the DCF77 signal, the beginning of each second can be easily recognized, providing a time accuracy in the order of $0.1$~s. Using an additional phase modulated signal superimposed on the amplitude modulation (AM) signal and using correlation receiver, the synchronization accuracy can be improved to a few hundreds of microseconds \cite{DCF_SDR}.
	\subsubsection{FM-RDS}
	\label{sec:FM-RDS}
	RDS is a supplementary digital data service that is superimposed on the regular transmission of an FM broadcast station. RDS is intended to provide additional, real-time services including, but not limited to, road traffic information. The digital data are transmitted via double-sideband AM on a suppressed $57$~kHz carrier, which is the third harmonic of the pilot tone for FM stereo transmission. Data are biphase coded, and the data rate is $1187.5$~bps. Data are hierarchically arranged into 16-bit words, 26-bit blocks, and 104-bit groups. 
	Depending on the data transmitted, \mbox{FM-RDS} defines multiple types of groups, each identified by a 4-bit group type code.
	A clock time and date (CT) group exists, purposely designed to transmit time information for receiver synchronization.
	A \mbox{CT-group} is transmitted once per minute, but the provided accuracy is poor, in the order of $100$~ms \cite{RDS}.
	Unfortunately, the CT may be absent in some RDS data stream, since it is not mandatory and some broadcasting stations do not use it. Moreover, using \mbox{FM-RDS} for absolute synchronization is unreliable, since broadcasting stations are not required to be UTC synchronized. Subsequently, when group transmissions are considered, relative time synchronization can be achieved with an error in the order of hundreds of microseconds \cite{RDSSynch}, however, offset with UTC may be unknown. Finally, it has to be stressed that low-power consumption \mbox{FM-RDS} receivers can be realized \cite{RDSConsumption}. 

	\section{The proposed communication mechanism} 
	\label{sec:ProposedApproach}
	In this work, the use of slotted ALOHA is suggested as a viable approach in dense LoRaWAN networks, under the name of \mbox{S-LoRa} protocol, to improve the overall energy efficiency and scalability. Within the scope of this work, we consider a monitoring application scenario with uplink only traffic. As stated before, in such a case the need for downlink frames for transferring time-related information can lead to excessive usage of communication opportunities, already bounded by severe constraints in terms of available bandwidth and duty-cycle limits. For this reason, differently from other works in the literature, out-of-band synchronization is used.
	A common sense of time among nodes is often required by the application to manage events of interests (e.g., in monitoring applications, to periodically sample the quantity of interest and schedule the transmission of the acquired values). In this work, a common sense of time is added to the original LoRaWAN stack, and used to define the timeslot boundaries within a contention window interval $T_{CW}$. In particular, relative time synchronization (in which a synchronization event simultaneously received by all the nodes is tracked) is considered. In detail, the synchronization procedure requires:
	\begin{itemize}
		\item two successive synchronization events, separated by a known interval, to estimate the local clock rate drift with respect to the common reference; 
		\item a single synchronization event for correcting the clock phase offset error.
	\end{itemize}
	All the technologies discussed in Sec. \ref{sec:Time_Diss_Tech} can be adopted; however, in this work, we focus on the \mbox{FM-RDS} solution that, although not offering the best accuracy, provides indoor coverage and ensures reliability due to the many FM stations broadcasting in a single region.
	
\begin{figure}[!htp]
		\centering
		\includegraphics[width=1\columnwidth]{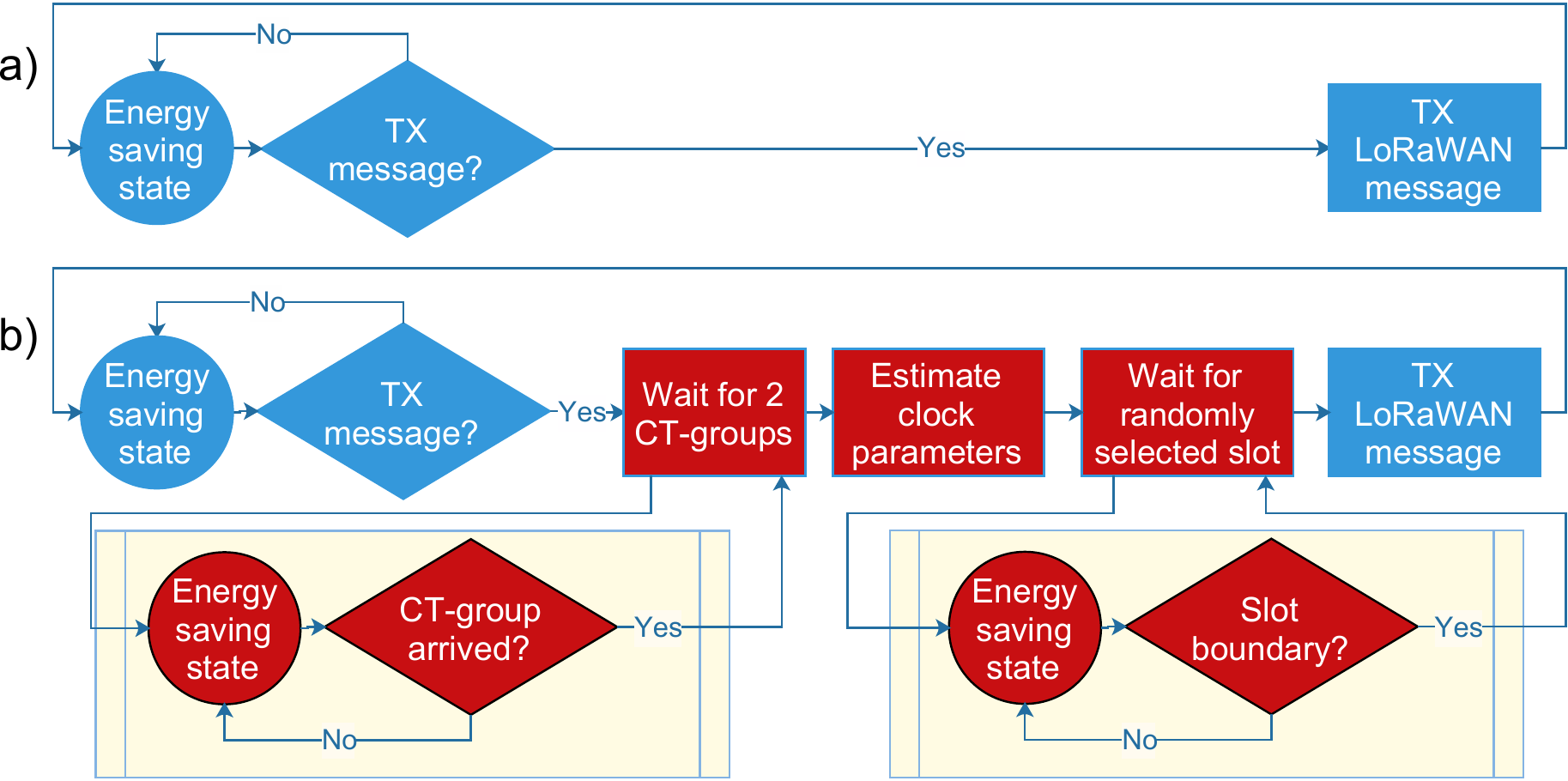}
		\caption{Flowchart of: a) regular LoRaWAN transmissions, b) the proposed \mbox{S-LoRa} approach based on out-of-band time dissemination.}
		\label{fig:flow_chart}
	\end{figure}

	As previously discussed in Sec. \ref{sec:FM-RDS}, \mbox{CT-groups} transmitted by an FM broadcaster contain the time information, but the accuracy is poor. 
	Consequently, we propose to use the \mbox{FM-RDS} \mbox{CT-group} detection as a common reference for the relative synchronization of the LoRa devices. The timestamp contained in each \mbox{CT-group} is only used during the initialization procedure (i.e., when a node joins the network). The only requirement is that all the nodes must simultaneously recognize the same \mbox{CT-group} and accurately timestamp it with the local clock.
	Fig.\ref{fig:flow_chart} shows a flowchart of the proposed transmission scheme, including the main steps required for a device to synchronize and transmit a message to the gateway, compared to a regular event-based LoRaWAN transmission.
	
\begin{figure}[!t]
		\centering
		\includegraphics[width=\columnwidth]{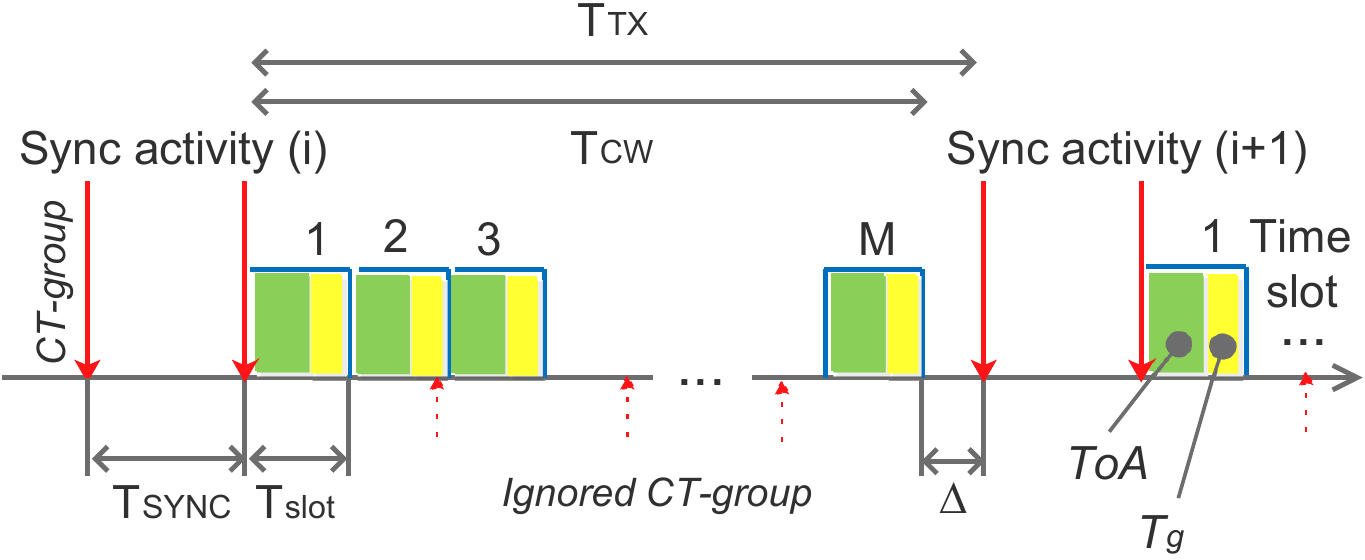}
		\caption{Structure of the timeslots for the proposed \mbox{S-LoRa} solution. Synchronization process requires the reception of two \mbox{CT-groups} by the \mbox{FM-RDS} radio.}
		\label{fig:frame}
		\vspace{-10pt}
	\end{figure}

	To start the synchronization process, a device wakes up its \mbox{FM-RDS} receiver in the proximity of the beginning of a contention window, compensating for the accumulated clock drift, and waits for a \mbox{CT-group}.
    After a $T_{\textrm{SYNC}}$, denoting the \mbox{CT-group} periodicity nominally equivalent to one minute, the node receives a second \mbox{CT-group}, which allows estimating the clock drift. The second \mbox{CT-group} also marks the beginning of the new contention window.
	Once the synchronization is completed, the device waits for a randomly chosen slot to transmit its message to the GW. The timeslot duration (see Fig.~\ref{fig:frame}) is designed to contain a message transmission, lasting a time on air (ToA), plus a guard interval ($T_g$), i.e., $T_{\textrm{slot}}=ToA+T_g$, for avoiding the collisions between transmissions in different slots.
	Let $T_{TX}=k \cdot T_{\textrm{SYNC}}$, for some value $k\in\mathbb{Z}^+$, be the time interval available for transmitting, the contention window has a duration $T_{CW}=T_{TX}-\Delta$, where, $\Delta$ takes into account the non integer number of $T_{\textrm{SYNC}}$ intervals (nominally lasting for one minute) encompassing the actual contention window, and $T_{TX}$ is sized accordingly to the target refresh period. Hence, the number of available slots is   $M=\frac{T_{CW}}{T_{\textrm{slot}}}=\frac{T_{TX}-\Delta}{ToA+T_g}$.

	A message is successfully transmitted to the GW if there are no intra- and inter-slot collisions. The former occurs when more than one device chooses the same slot for transmission; the latter occurs when messages transmitted in adjacent timeslots partially overlap due to synchronization errors and transmission time uncertainties. Although not within the scope of this paper, it is possible to derive the optimal guard time $T_g$, maximizing the overall throughput. 
	 
	Finally, it is worth noticing that, despite the MAC mechanism of the devices has been modified, no LoRaWAN rule is violated and the message payload is unaltered. Accordingly, the proposed approach does not require changes in the legacy LoRaWAN backend infrastructure, ensuring low installation and maintenance costs. 

	\section{Timing errors}
	\label{TimingErrors}
	For any synchronization mechanism, timing errors arise from the propagation delay of the messages, and the non-ideal behavior of the transceivers and the local clocks. In this section, the factors  contributing to the uncertainty of the transmission time of a device in the proposed communication mechanism are discussed and characterized.
	
	\subsection{Transmission timing error}
	A transceiver is responsible for transmitting the LoRa frame once the timeslot deadline has occurred. Unfortunately, traveling across the lower layers of the node's protocol stack adds delays, which cannot be easily entirely removed even by proper calibration. Typically, such delays depend on the software/hardware being used, and can be different from manufacturer to manufacturer. Therefore, the Type A uncertainty $u_{TX}$ must be taken into account, which we evaluated experimentally (see Sec.~\ref{sec:exp_Tx_latency}). 
	
	\subsection{Message propagation delay}
	Two messages transmitted precisely at the start of their respective timeslots can still collide at the receiver due to the propagation delay. 
	In the case of line-of-sight conditions, the propagation delay increases linearly with the distance of a device from the gateway. Assuming that the devices are uniformly distributed in an annular shaped region, with radii $R_L$ and $R_l$, the probability distribution $f_{PD}(x)$ of the propagation delay for a device is given by 
	\begin{equation}\label{fpd_x}
	\begin{aligned}
	f_{PD}\left(x\right) =&\begin{cases}
	\frac{2xv^2}{R_L^2-R_l^2} & \text{if } \frac{R_l}{v}\leq x\leq \frac{R_L}{v}\text{,}\\
	0  & \text{otherwise}\text{,}
	\end{cases}
	\end{aligned}
	\end{equation}
	where $v$ is the speed of light.
	The distribution $f_{PD}(x)$ is a triangular probability density function with mean $\mu_{PD}$ and standard deviation $\sigma_{PD}$ 
	\begin{equation}
	\begin{aligned}
	\mu_{PD}=\frac{2\left(R_L^2+R_l^2+R_LR_l\right)}{3\left(R_L+R_l\right)v} \end{aligned}
	\end{equation}
	\begin{equation}
	\begin{aligned}
	\sigma_{PD}=\frac{\left(R_L-R_l\right)\sqrt{R_L^2+R_l^2+4R_LR_l}}{3\sqrt{2}v\left(R_L+R_l\right)}.
	\end{aligned}
	\end{equation}
	For devices distributed on a disk-shaped region of radius $R$, we have $R_L=R$ and $R_l=0$, giving $\mu_{PD}=\frac{2}{3}\frac{R}{v}$ and  $\sigma_{PD}=\frac{R}{3\sqrt{2}v}$.
	The Type B uncertainty $u_{PD}$ can be estimated as 
	\begin{equation}\label{eq:u_PD}
	u_{PD}=\sqrt{\mu_{PD}^2+\sigma_{PD}^2},
	\end{equation}
	the average value $\mu_{PD}$ is taken into account since it is not a-priori known.
	
	\subsection{Local clock error}
	\label{sec:clk_error}
	The natural or true clock is generally denoted by $t$ and runs at a rate of 1 second per second. The origin $t=0$ occurs at some arbitrary instant. Unfortunately, real-world implementations of local clocks are imperfect, hence their internal value is represented by the quantity $C(t)$ when evaluated at the true instant $t$. Typically, a device clock is implemented through a timer peripheral in a microprocessor-based system. Assuming the timer is running at the nominal rate $F_0$, the timer period $T_0=1/F_0$ is the resolution of $C(t)$, i.e., the smallest unit by which it is updated. Accordingly, when a first-order clock model is adopted (sometimes referred to as the SKM model \cite{SKM}), $C(t)$ is described as
	\begin{equation}\label{ct}
	C(t) \approx (\alpha+\beta t)T_0,
	\end{equation}
	where the quantity $\alpha\cdot T_0$ is the time difference between local clock time and true time at the origin, and $\beta\cdot T_0$ is the real local clock rate. The clock drift $\gamma$ and offset $\theta (t)$ can be, respectively, defined as
	\begin{equation}\label{drift}
	\gamma =\frac{dC(t)}{dt}-F_0T_0=\frac{dC(t)}{dt}-1=(\beta-F_0)T_0,
	\end{equation}
	\begin{equation}\label{offset}
	\theta (t) = C(t)-t.
	\end{equation}
	For a perfectly syntonized clock, the drift (or skew) is $\gamma =0$ and $\beta =F_0$; for a perfectly synchronized clock the offset is $\theta (t)=0$ and $\alpha =0$.
	
	As a consequence, from the node point of view, the start of the next transmission (not necessarily the start of the next timeslot), $t_d$ in the true time reference, is seen as the local clock value $C(t_d)$. If a synchronization procedure is carried out by the node, it is possible to assume that the timer is restarted and the actual clock value is represented by
	\begin{equation}\label{actualct}
	V(t_d)=C(t_d)-C(t_0)=\beta (t_d-t_0)T_0\text{,}
	\end{equation}
	where $t_0$ is the true clock value at the origin of the current synchronization interval.
	
	Thus, the timing error of a node relative to the start of a message transmission is the result of the three different contributing factors: the uncertainty in the syntonization procedure ($u_\beta$) causing the transmitter to drift from the nominal rate $F_0$; the uncertainty $u_{t_0}$ in the $t_0$ instant (e.g., due to the detection mechanism of the synchronization event); and the uncertainty $u_{t_d}$ in the transmission time $t_d$.
	The overall uncertainty $u_V$ can be obtained from \eqref{actualct} by propagating these contributions, resulting in
	\begin{equation}\label{overalluncertainty}
	u_V=\sqrt{\left(\frac{\partial {V}}{\partial{\beta}}\right)^2\!\cdot u_\beta^2  + \left(\frac{\partial {V}}{\partial{t_d}}\right)^2\!\cdot u_{t_d}^2 + \left(\frac{\partial {V}}{\partial{t_0}}\right)^2\!\cdot u_{t_0}^2}\text{,}
	\end{equation}
	which is finally evaluated as 
	\begin{equation}\label{uvfinal}
	u_V=T_0\sqrt{\left(t_d-t_0\right)^2\cdot u_\beta^2  + \beta ^2\cdot u_{t_d}^2 + \beta ^2\cdot u_{t_0}^2}\text{.}
	\end{equation}
	
\begin{figure}
		\centering	
		\includegraphics[width=0.9\linewidth]{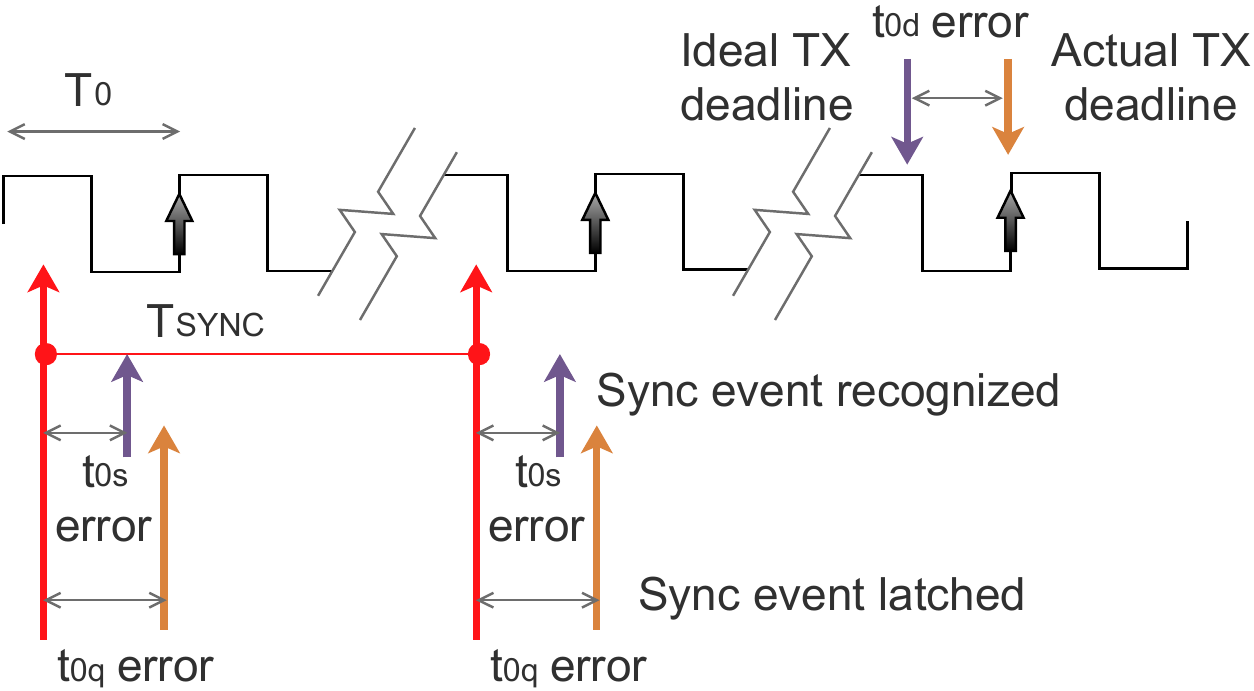}\hfill
		\caption{Timing errors diagram highlighting the contributions of $u_{t_0s}$ and $u_{t_0q}$ to the clock synchronization uncertainty, and $u_{t_0d}$ to the message transmission uncertainty.}
		\label{fig:timing_errors}
		\vspace{-10pt}
	\end{figure}

	It is possible to further elaborate the uncertainty contributions using Fig.~\ref{fig:timing_errors}, in which it is assumed that the devices can latch the synchronization events (e.g., the reception of a new \mbox{CT-group}) and transmit a new message only on the rising edge of the local clock $V(t)$ running at $F_0$.
	Accordingly, the actual deadline $\hat{t_d}$ is affected by the time quantization error uniformly distributed between 0 and $T_0$, resulting in the uncertainty $u_{t_d}$ as 
	\begin{equation}\label{utd}
	u_{t_d}=\frac{T_0}{\sqrt{3}}.
	\end{equation}
	The actual reference time instant $\hat{t_0}$ is affected by two errors, i) the time quantization, as in \eqref{utd}, and ii) the timing error in the detection of the synchronization event. These errors are characterized by the uncertainties $u_{t_0q}$ and $u_{t_0s}$, respectively, leading to
	\begin{equation}\label{ut0}
	u_{t_0}=\sqrt{u_{t_0q}^2+u_{t_0s}^2}=\sqrt{\frac{T_0}{3}^2+u_{t_0s}^2}.
	\end{equation}
	
	Finally, the syntonization mechanism must be analyzed for estimating the uncertainty $u_\beta$.
	
	The $T_{\textrm{SYNC}}$ comprises the interval between two \mbox{CT-group} transmission, i.e., $T_{\textrm{SYNC}}=60$~s. Consequently, the rate $\beta$ can be estimated as
	\begin{equation}\label{bhat}
	\hat{\beta}=\frac{\Delta_{\textrm{COUNTER}}}{T_{\textrm{SYNC}}},    
	\end{equation}
	where $\Delta_{\textrm{COUNTER}}$ is a value representing the number of clock periods $T_0$ counted in $T_{\textrm{SYNC}}$. Using \eqref{bhat} to propagate the uncertainty of $T_{\textrm{SYNC}}$ (i.e., $u_{T_{\textrm{SYNC}}}$) leads to
	\begin{equation}\label{ubeta}
	u_\beta = \frac{\Delta_{\textrm{COUNTER}}}{T_{\textrm{SYNC}^2}}\cdot u_{T_{\textrm{SYNC}}}=\frac{\beta}{T_{\textrm{SYNC}}}\cdot u_{T_{\textrm{SYNC}}}.
	\end{equation}
	The estimation $\hat{T}_{\textrm{SYNC}}$ of the synchronization interval is affected by the time quantization, as well as the error in the detection of the reference time signal. Differently from $\hat{t_0}$, their effects occur both at the start and at the end of the interval itself. Consequently, the uncertainty is
	\begin{equation}\label{usync}{
		u_{T_{\textrm{SYNC}}}=\sqrt{4u_{t_0q}^2+4u_{t_0s}^2}=\sqrt{4\frac{T_0}{3}^2+4u_{t_0s}^2}.}
	\end{equation}
	When the worst case scenario is considered, $t_d-t_0 \simeq (k \cdot T_{\textrm{SYNC}}) = T_{TX}$ for some $k$, so that we can simplify $u_V$ as 
	\begin{equation}\label{uvworst}
	\begin{aligned}
	u_V & =\beta T_0\sqrt{k^2 \cdot u_{T_{\textrm{SYNC}}}^2 + u_{t_d}^2 + u_{t_0}^2}\\
	& =(1+\gamma)\sqrt{(4k^2+2)\frac{T_0}{3}^2+(4k^2+1)u_{t_0s}^2}.
	\end{aligned}
	\end{equation}
	Assuming $k>>1$, $T_0<<u_{t_0s}$ and $\gamma<<1$, as typically occurs when synchronization via \mbox{FM-RDS} signal is considered, \eqref{uvworstsimplify} reduces to
	\begin{equation}\label{uvworstsimplify}
	\begin{aligned}{
		u_V=2k \cdot u_{t_0s}.}
	\end{aligned}
	\end{equation}
	
	Although $T_0$ is generally known, $u_{t_0s}$ must be evaluated, e.g., by performing experiments in agreement with Type~A uncertainty estimation. Such an approach is adopted in this work, as detailed in Sec.~\ref{Experimental Results}.
	\subsection{Concluding remarks}
	\label{sec:remarks_uncertainty}
	All the previously addressed errors contribute to the overall timing error introduced in this section. Without loss of generality, the errors can be grouped into the overall timing uncertainty $u$ given by 
	\begin{equation}\label{u}
	u=\sqrt{u_{TX}^2 + u_{PD}^2 + u_{V}^2}.
	\end{equation}
	The value of $u$ can be transformed into an equivalent standard deviation of a probability density function describing the possible timing error population and used in the simulations of the considered application scenario.
	It has to be emphasized that the proposed approach relies on the  simultaneous estimation of the local clock rate by all the devices using the same synchronization events, so that variations in $T_{\textrm{SYNC}}$ affect all devices, without compromising the timeslot boundaries.
	
	\section{Experimental results}
	\label{Experimental Results}
	The accurate characterization of IoT components is known to be a main concern of current research activities, requiring specific instrumentation setup and measurement methods \cite{IoTMeas}. In this section, we describe the experimental validation of hardware-assisted synchronization required for implementing the \mbox{S-LoRa} protocol. We purposely designed an experimental setup with twofold objectives:
	\begin{itemize}
		\item to evaluate the uncertainty in transmitting a LoRaWAN message in a well-defined time instant as needed by a timeslot deadline,
		\item to evaluate the uncertainty in retrieving time information from the \mbox{FM-RDS} broadcast signal.
	\end{itemize}
	In the considered scenario, focused on local relative time synchronization, the propagation delay is generally negligible, given that devices are a few kilometers apart from the gateway and the $u_{PD}$ is in the order of a few microseconds, no matter the adopted SF.
	Each network node uses daughter radio cards, which are connected to a host system (LoRa via an SPI link, and FM via I2C link), executing the upper layers of the protocol stack. The LoRa radio is based on embedded SX1272\footnote{Datasheet available at www.semtech.com. See also \cite{LoRaTool}.} daughter board, which is capable of signaling the completed transmission (TX\_DONE) or reception (RX\_DONE) of LoRa frames by a digital line on the board connector.
	This line is used to characterize the LoRa transceiver without the additional latency caused by the execution of the protocol stack on the host system.
	
	Regarding the \mbox{FM-RDS} decoder, the Si4703 from Silicon Labs{\footnote{Datasheet available at www.silabs.com}} is used, hosted in an FM Radio Tuner Evaluation Breakout Board from Hitletgo. This device implements a complete solution for decoding both the audio and RDS signals, thus including many unused blocks (e.g., the driver for the headphones); accordingly, it is not possible to evaluate precisely the power consumption. When the "verbose" mode is active, a hardware pulse is generated each time a new RDS group is received. This hardware line represents the actual time reference with respect to the broadcasting station.
	The host board is an STM32 Nucleo board, running the upper layers of the protocol stack (other than analyzing the content of the incoming RDS \mbox{CT-group}).
	
	\subsection{Characterization of LoRa frame transmission latency}
	\label{sec:exp_Tx_latency}
\begin{figure}[!ht]
		\centering
		\includegraphics[width=\columnwidth]{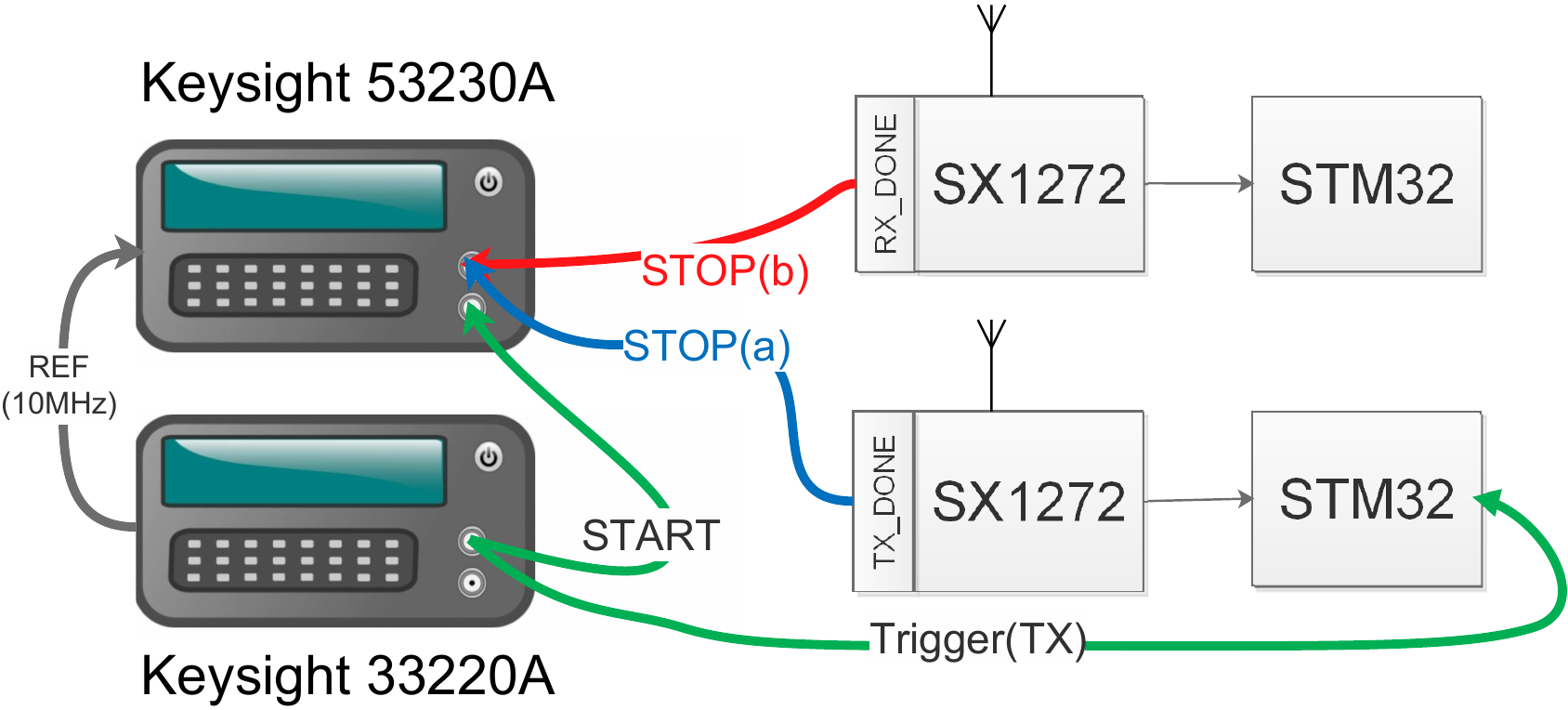}
		\caption{Testbed for evaluating LoRa frame transmission latency.}
		\label{fig:LoRaTX}
		\vspace{-10pt}
	\end{figure}

	We used two nodes (A and B) to characterize the capability of the LoRa radio to timely transmit a new frame.
	In Node A, the Nucleo board is configured to start the transmission after the recognition of a rising edge on its input line associated with the highest interrupt priority. The trigger signal was provided by a function generator (Keysight 33220A).  To ensure a deterministic behavior and minimize the overall latency, the frame content was already available in memory before the trigger occurred. The Node B is permanently configured in the reception mode to detect the LoRa frame.
	A counter (Keysight 53230A, receiving the $10$~MHz reference signal from the function generator) is used (see Fig.~\ref{fig:LoRaTX}) for measuring the time $\Delta_{TX}$ elapsing between the trigger signal edge and the TX\_DONE edge in Node A (configuration-a), and between the trigger signal edge and the RX\_DONE edge in the Node B (configuration-b). 
	The $\Delta_{TX}$ mean value reflects message duration and hardware fixed delay (that can be removed by proper calibration), while the standard deviation is $\sigma_{\Delta_{TX},a}=1.4~\mu$s in configuration-a and $\sigma_{\Delta_{TX},b}=2.8~\mu$s in configuration-b, when uncertainties of both Node A and Node B are combined; such values stay the same for different SFs. In this way, it has been possible to estimate the uncertainty in transmitting a LoRa frame $u_{TX}=1.4~\mu$s. For the sake of completeness, note that the resolution of instantaneous frequency jumps inside a single LoRa symbol is $T_{jump} = T_C/2^{\textrm{SF}}$, which is about $8~\mu$s for $\textrm{SF}=7$. 
	
	\subsection{Characterization of \mbox{FM-RDS} \mbox{CT-group} reception}
\begin{figure}[!htp]
		\centering
		\includegraphics[width=0.8\linewidth]{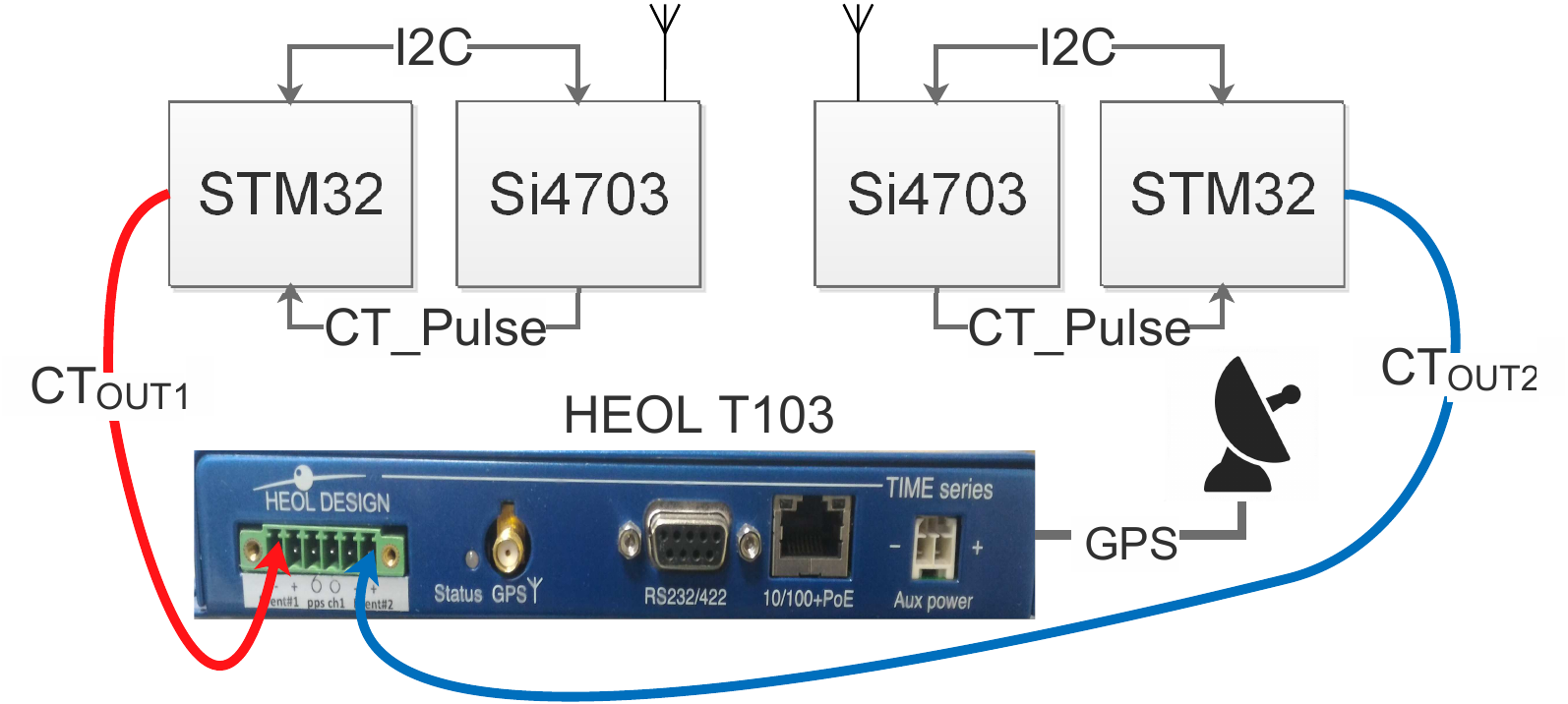}
		\caption{Testbed for evaluating \mbox{CT-group} message latency.}
		\label{fig:CT_Si4703}
		\vspace{-10pt}
	\end{figure}

	Two (co-located) nodes are used to characterize the capability of the FM receivers to detect an incoming \mbox{CT-group}. Each node follows the \mbox{FM-RDS} data stream continuously looking for \mbox{CT-groups}. Once a new \mbox{CT-group} is detected, a pulse is generated on an STM32 digital output line ($CT_{OUT1,2}$ in Fig.~\ref{fig:CT_Si4703}). The digital lines are acquired by a GPS-synchronized time server T103 from HEOL, which is able to timestamp the two line edges with sub-microsecond accuracy with respect to the UTC time. 
	Initially, the population $T_{CT_{OUT1}}$ of $CT_{OUT1}$ timestamps is considered for evaluating the jitter in the \mbox{CT-group} periodicity, i.e., for evaluating the uncertainty $u_{CT}$ in the time interval (nominally equal to $T_{\textrm{SYNC}}$) elapsing between two successive \mbox{CT-groups}. In particular, we consider the standard deviation $\sigma_{CT_{OUT1}}=u_{CT}$. Recall that $u_{CT}$ does not affect the proposed synchronization strategy.
	Subsequently, the time difference of arrival of the pulse at the nodes' digital output lines $\Delta _{CT}=T_{CT_{OUT1}}-T_{CT_{OUT2}}$ is considered for taking into account the uncertainties of both nodes, which ideally would be equal. Because of systematic errors, summarized by the average value $\mu_{\Delta_{CT}}$, the \mbox{CT-group} reception uncertainty is estimated as  
	
	\begin{equation}\label{eq:ut0s}
	\begin{aligned}
	u_{t_0s}= \sqrt{\frac{\mu_{\Delta_{CT}}^{2}+\sigma_{\Delta _{CT}}^{2}}{2}},
	\end{aligned}
	\end{equation}
	 where $\sigma_{\Delta_{CT}}$ is the standard deviation of the $\Delta_{CT}$ population.

	The value of $u_{CT} $ depends primarily on the mechanism employed by the FM broadcaster for the transmission of \mbox{CT-groups} (in agreement with \mbox{FM-RDS} specifications) and for the synchronization  with the UTC absolute time. A two-day long measurement campaign was carried out, syntonizing both the Si4703 devices on three different Italian FM broadcasting radio-stations. The ``Radio Monte Carlo" station achieved a $u_{CT} = 171$~ms, the ``RTL102.5" station showed a $u_{CT} = 62$~ms, and the ``Radio Freccia" obtained a $u_{CT} = 114$~ms. Only, the best result is considered for the following part of the paper, thus $u_{CT} = 62$~ms is used in the simulation of Sec.~\ref{Simulator Results}. Additionally, for ``RTL102.5", $\mu_{\Delta_{CT}}=0.41$~ms and $\sigma_{\Delta_{CT}}=0.24 $~ms, so that according to our experiments $u_{t_0s}=0.34$~ms.
	
	\section{Simulator Results}
	\label{Simulator Results}
	To evaluate the performance of the proposed slotted communication based on out-of-band synchronization for a wide area metering application, we resorted to using simulations. Indeed, simulations permit to  easily and effectively evaluate networks with thousands of devices. The performance is characterized in terms of transmission success probability and energy efficiency, the latter being the primary concern for any IoT wireless application.

	\subsection{The simulator}
	{In this work, to simulate the proposed communication mechanism, we modified LoRaEnergySim~\cite{8885739} Python simulator to incorporate the experimental and analytical model results presented in Sec.~\ref{TimingErrors} and Sec.~\ref{Experimental Results}\footnote{Simulator code available at https://github.com/Beltra90/LoRaMACSim}. Although numerous LoRa/LoRaWAN simulators have been designed and reported in the literature (for detailed comparison see e.g., \cite{haxhibeqiri2018survey}), careful consideration of the power capture effect in each simulator is required for evaluating any new proposals. In general, these simulators are based on two alternative power capture models,  either cumulative or dominant interference. For scenarios with many interfering devices, the dominant interference model adopted by LoRaEnergySim provides an upper bound to the performance of LoRa compared to the cumulative interference model proposed in~\cite{7996384}. Consequently,  when it is used to evaluate the performance improvements of the proposed slotted communication, which intrinsically reduces the number of interfering signals compared to the non-slotted LoRaWAN, a dominant interference model provides more conservative results. The salient features of the interference model of LoRaEnergySim are as follows}:
	a) signals transmitted using different SFs are orthogonal, b) messages can only be correctly received if they satisfy the minimum signal-to-interference ratio (SIR) and signal-to-noise ratio (SNR) thresholds, c) when computing the SIR, only the dominant interferer is considered, and d) messages can potentially survive a partial overlap limited to the first few symbols of the preamble. 
	To model the time uncertainty affecting a message transmission, we opted to use the worst-case combined uncertainty $u$ determined in Sec.~\ref{sec:clk_error}.
	Each transmission is affected by an independent and identically distributed stochastic offset. The standard deviation of the timing offset distribution affecting a transmission is chosen to be equal to the overall timing uncertainty found in Sec.~\ref{sec:remarks_uncertainty}, \eqref{u}.
	This model of the timing errors via the combined uncertainty allowed us to simulate scenarios with a large number of devices, avoiding the computationally expensive task of individually modeling the local clock of each device. 
	
	\subsection{Simulated scenarios}
	We simulated a typical urban deployment scenario in which a single GW is located at the center of a group of devices. The devices are randomly and uniformly distributed over a circular-shaped region with a diameter of $3$~km. The signal propagation in the urban scenario was modeled similarly to~\cite{7996384}, however, we opted to use an uncorrelated lognormal shadowing fading with standard deviation of $7.8$~dB~\cite{7377400}. The parameters used in the simulation are reported in Table~\ref{tb:par}. A convenient, linearly increasing, guard time $k \cdot T_{g0}$ (where $k$ is the number of $T_{\textrm{SYNC}}$ in an $T_{TX}$) was selected to maximize the throughput given the synchronization uncertainty $u$. We tested different configuration of device densities, SFs, payload sizes and transmission intervals $T_{TX}$, always respecting the limitations imposed by the LoRaWAN specifications. The traffic in the scenario consists of unconfirmed uplink messages, i.e., from end devices to the GW. Each device generates a message once every $T_{TX}$, and the cumulative number of messages transmitted by all devices during each simulation run is $200 000$. Consequently, in each run, the network is simulated for a time duration ranging from one to 90 hours. For each combination of SF, $T_{TX}$, and payload length $L_P$, the results are averaged over ten simulation runs.
	The complete list of parameters used in the simulation is available in Table~\ref{tb:par}.
	
	\begin{table} [!htb]
		\caption{LoRa Parameters used for the performance evaluation}
		\centering
		\scalebox{0.97}{
			\begin{tabular}{ | l | l | l | }
				\hline 
				\textbf{Parameter} &\textbf{Symbol} & \textbf{Value} \\\hline\hline
				Bandwidth and Channel& BW and CH & $125$~kHz and $868.1$~MHz\\
				Preamble  &  & $8$ bytes \\
				Coding Rate &  CR &$4/8$  \\     
				Spreading Factor  & SF & $\in \left\{7,9,12\right\}$\\
				Payload @ SF7  &$L_{P7}$ & $\in \left\{10,51,221\right\}$ bytes \\ 
				Payload @ SF9  &$L_{P9}$ & $\in \left\{10,51,115\right\}$ bytes \\ 
				Payload @ SF12  &$L_{P12}$ & $\in \left\{10,51\right\}$ bytes \\ 
				Guard time &$T_{g0}$ & $3$~ms\\
				Noise Figure and PSD & & $6$~dB and $-174$~dBm/Hz\\
				Shadow fading& &$7.8$~dB\\
				GW Antenna Height& &$15$~m\\
				Device floor no.&&$\mathcal{U}(1,4)$\\
				Internal walls no.&&$\mathcal{U}(0,3)$\\
				SNR Thresholds& &$\left[-6,-12,-20\right]$~dB\\
				SIR Threshold& &$1$~dB\\
				Transmitting Power& & $14$~dBm\\
				\mbox{CT-group} duration&$T_{CT}$ & $86.7$~ms\\
				Radio switch on time&$T_{ON}$ & $1$~ms\\
				\hline
		\end{tabular}}
		\label{tb:par}
		\vspace{-10pt}
	\end{table}
	\subsection{Results: success probability and energy efficiency}
\begin{figure*}[!htb]%
	\centering\captionsetup{justification=centering}
	\subfloat[$N=5000$ nodes in the network ]{\includegraphics[width=0.45\linewidth]{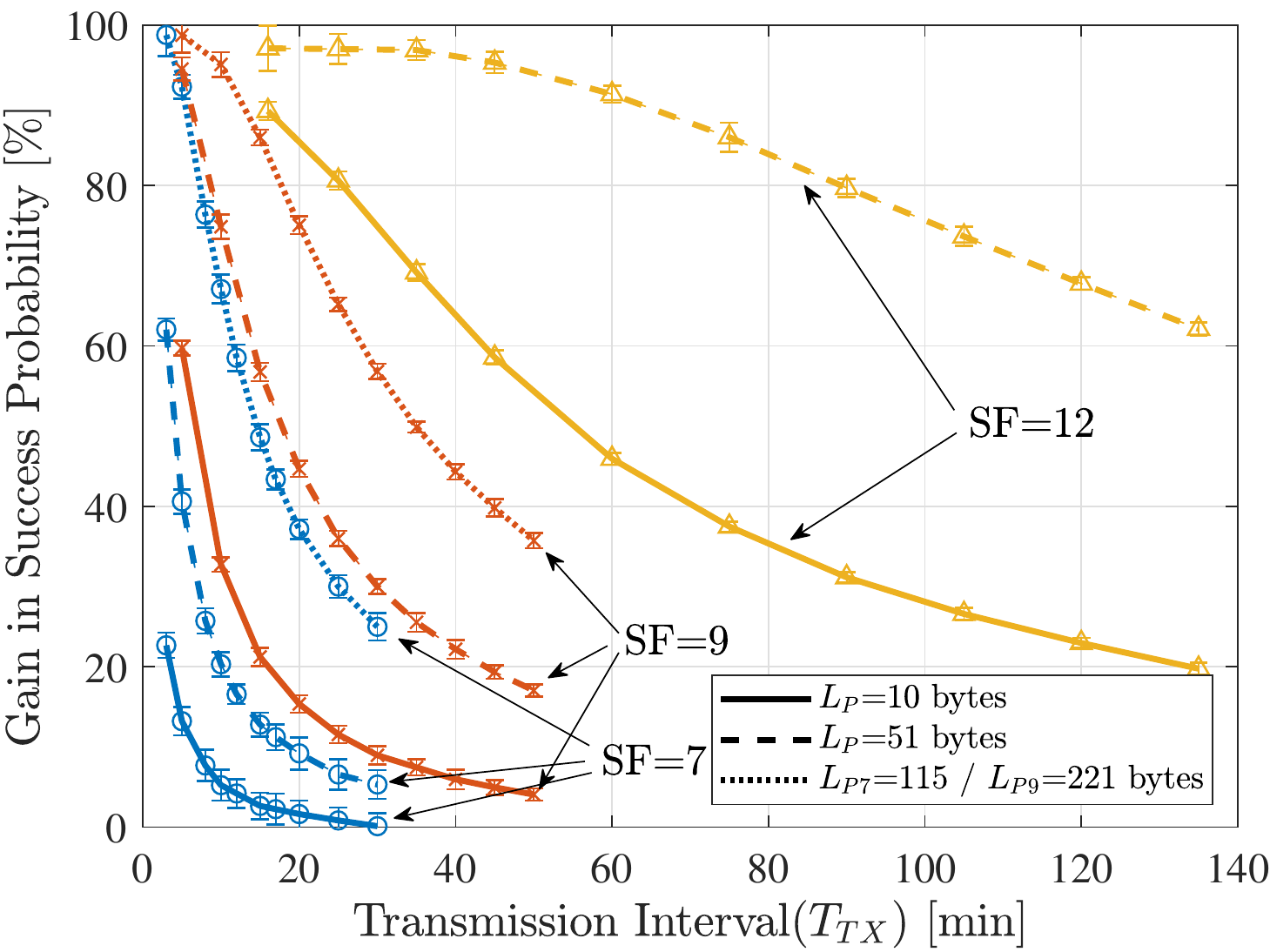}}\hspace{45pt}
	\subfloat[$N=10000$ nodes in the network]{\includegraphics[width=0.45\linewidth]{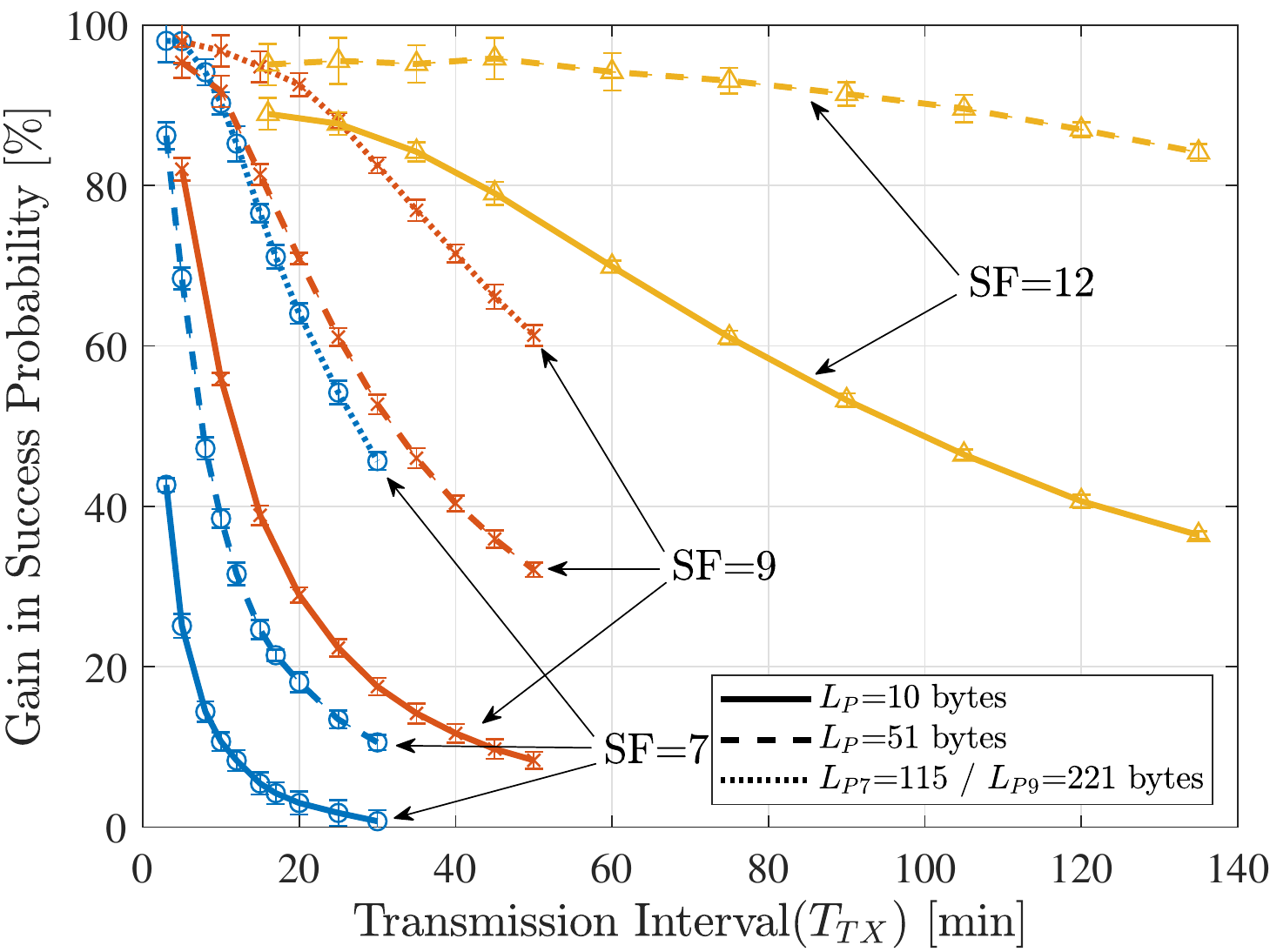}}
	\caption{Percentage relative success probability gain of \mbox{S-LoRa} over LoRaWAN.}
	\label{fig:sp}
	\vspace{-10pt}
    \end{figure*}

	The propoposed \mbox{S-LoRa} has been compared against the pure ALOHA-based transmission of a LoRaWAN Class-A device transmitting unconfirmed messages. We selected the (transmission) success probability as first metric of interest, calculated as the fraction of generated messages that are successfully received by the GW. For the set of parameters under investigation, the success probability of \mbox{S-LoRa} and LoRaWAN was always observed to be smaller than 0.5. It is possible to achieve significantly higher values by decreasing the average distance of the devices from the gateway.
	However, this type of consideration is beyond the scope of this paper, which instead primarily focuses on the relative comparison of the performance of LoRaWAN and \mbox{S-LoRa}.  For this reason, in Fig.~\ref{fig:sp} the relative success probability gain of \mbox{S-LoRa} with respect to the reference LoRaWAN ALOHA-like implementation is shown, for a network consisting of $N=5000$ (Fig.~\ref{fig:sp}.a), and  $N=10000$ (Fig.~\ref{fig:sp}.b) devices.  
    Depending on the transmission interval, ToA, and SF considered in the scenario, improvements close to 100\% in the success ratio can be reached, confirming the feasibility and advantages of the proposed approach. 
    As the scenario becomes increasingly demanding, i.e., short $T_{TX}$ and large $L_P$, the gain in success probability of \mbox{S-LoRa} over LoRaWAN becomes more significant.
	
	
	The other performance metric considered by this study is the energy efficiency. It is important to investigate the energy efficiency of \mbox{S-LoRa} to determine if the increased energy consumption resulting from the synchronization mechanism is adequately compensated by an improvement in the communication performance metric (i.e., the transmission success probability).  
	In calculating the energy consumption and energy efficiency the following assumptions have been made: 
	\begin{itemize}
		\item both regular LoRaWAN and \mbox{S-LoRa} require an always-on clock for scheduling monitoring activity and subsequent message transmission; therefore, the power consumption of the clock oscillator circuit was not taken into account in the comparison,
		
		\item each time a device wants to synchronize, two successive \mbox{CT-groups} must be received. The \mbox{FM-RDS} receiver remains active for $T_{RX1,\textrm{FM-RDS}} = T_{ON} + T_{CT} + T_{TX}\cdot\gamma + \frac{3}{\sqrt{2}}\cdot u_{CT}$  in order to receive the first synchronization event, where $T_{ON}$ and $T_{CT}$ are defined in Table~\ref{tb:par}. It is assumed that when the \mbox{FM-RDS} receiver is activated after a long time the device will account for the uncertainty of its clock drift. To receive the second \mbox{CT-group}, the \mbox{FM-RDS} remains active for $T_{RX2,\textrm{FM-RDS}} = T_{ON} + T_{CT} + 3 \cdot u_{CT}$, since in this case the device has adjusted its clock after the reception of the previous \mbox{CT-group}.
	\end{itemize}
	
    
    \begin{figure*}[!htb]%
	\centering
	\subfloat[$N=5000$ nodes in the network ]{\includegraphics[width=0.45\linewidth]{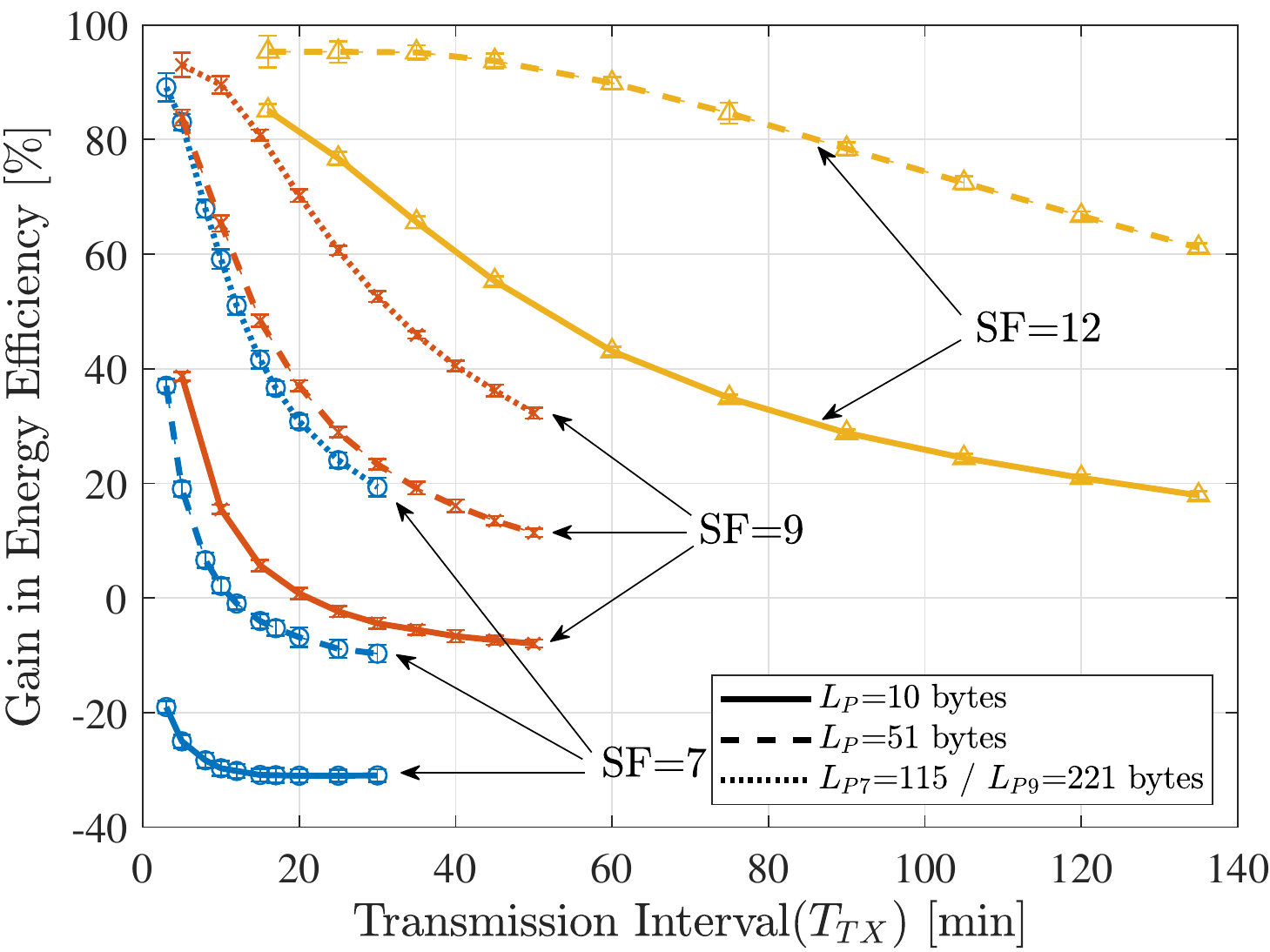}}\hspace{45pt}
	\subfloat[$N=10000$ nodes in the network]{\includegraphics[width=0.45\linewidth]{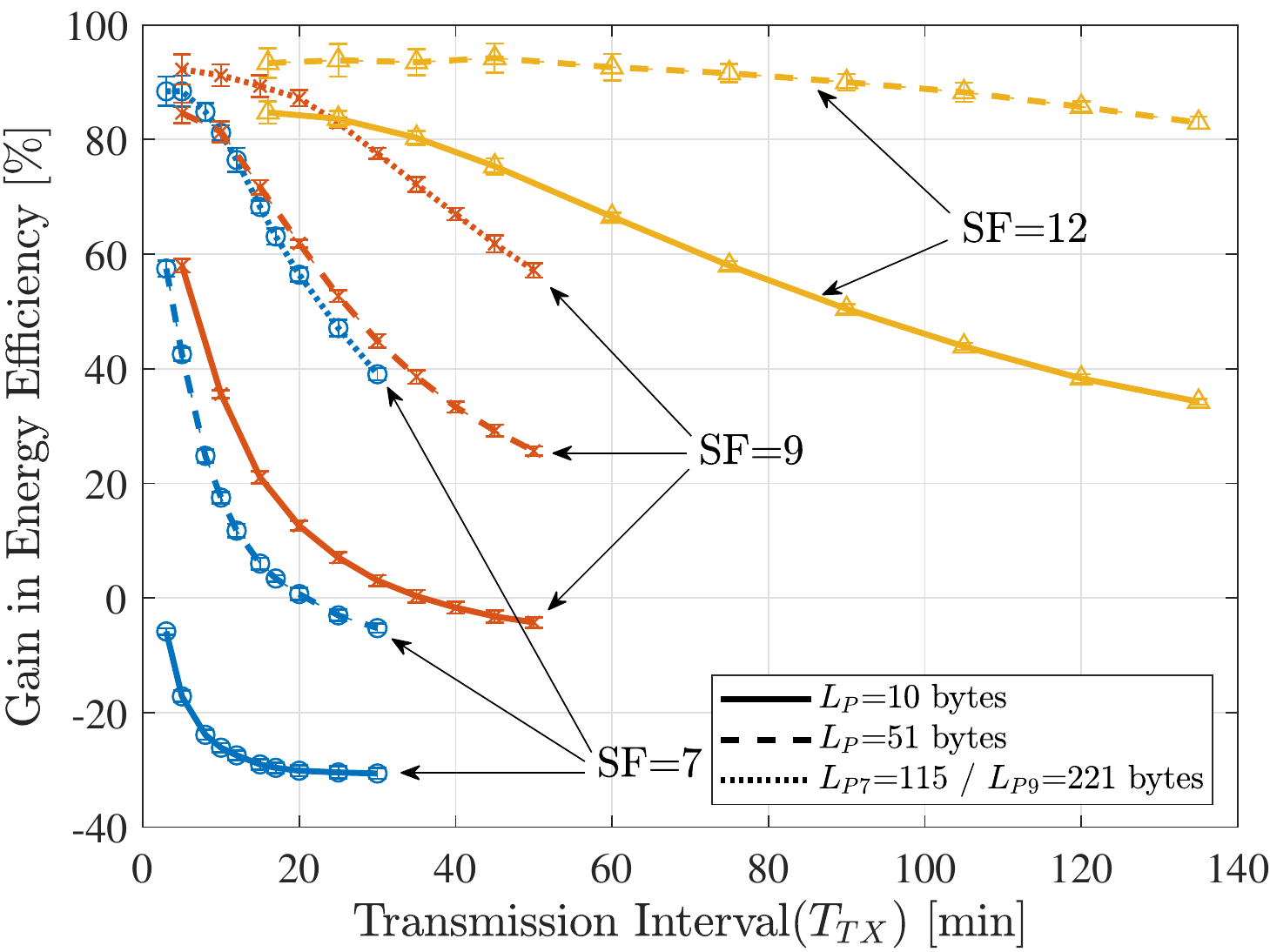}}
	\caption{Percentage relative energy efficiency gain of \mbox{S-LoRa} over LoRaWAN.}
	\label{fig:ef}
	\vspace{-10pt}
    \end{figure*}
	The LoRa Model Calculator Tool from Semtech \cite{LoRaTool} is used to derive the transmission and sleep currents for a SX1272/73 LoRa transceiver. Since downlink traffic is not considered (only unconfirmed uplink messages are sent), it is assumed that the LoRa transceiver is either in transmitting or sleeping state and never in receiving state. The current in transmission state is $I_{TX, \textrm{LoRa}} = 44$~mA, whereas the current during sleep is $I_{\textrm{sleep},\textrm{LoRa}} = 100$~nA. Accordingly, assuming a supply voltage $V_\text{dd} = 3.3$~V, the power consumption during message transmission is $P_{TX,\textrm{LoRa}} = 145.2$~mW, which decreases to $P_{\textrm{sleep},\textrm{LoRa}} = 0.33~\mu$W in sleep state.
	Unfortunately, \mbox{FM-RDS} receivers available on the market generally embed many functionalities other than RDS stream demodulation, thus the current consumption reported in many data sheets is significantly larger than the current strictly required for the demodulation of the RDS stream.  We used~\cite{RDSConsumption} as reference for the \mbox{FM-RDS} receiver's current consumption while active, where it is reported that the absorbed current is $I_{RX,\textrm{FM-RDS}} = 1.2$~mA with a supply voltage $V_\text{dd} = 3.3$~V, so that the power consumption during message transmission is $P_{RX,\textrm{FM-RDS}} = 3.96$~mW. When the \mbox{FM-RDS} receiver is not listening for \mbox{CT-groups}, it is assumed to be in idle mode; due to unavailability of current consumption in idle state in the literature, we consider instead a state-of-the-art device as the CC1125 Narrowband Transceiver \cite{tiCC1125}, with a current consumption of $I_{\textrm{idle},\textrm{FM-RDS}} = 120$~nA and a supply voltage of $V_\text{dd} = 3.3$~V, so that the power consumption in the idle state is $P_{\textrm{sleep},\textrm{FM-RDS}} = 0.36~\mu $W. We obtained the switching times between different radio states of the \mbox{FM-RDS} receiver from the same literature source, which we aggregated and upper bounded by the quantity $T_{ON}$.
	
	Relying on the previously described real-world hardware-related figures of merit, the energy efficiency of a device is calculated as the ratio of the overall number of bits successfully transmitted over the total energy consumption of the device during the entire simulation. In Fig.~\ref{fig:ef}, the relative energy efficiency gain with respect to the reference LoRaWAN implementation is shown, when the considered network consists of $N=5000$ (Fig.~\ref{fig:ef}.a), and $N=10000$ (Fig.~\ref{fig:ef}.b) devices. 
	Like for the success probability gain, we can observe that the gain in energy efficiency of \mbox{S-LoRa} over LoRaWAN becomes more significant for short transmission intervals and large payload sizes. We can also observe that although \mbox{S-LoRa} offers a strictly positive gain in the success probability over LoRaWAN, the energy efficiency gain can be negative (i.e., using \mbox{S-LoRa} can lead to a loss in energy efficiency).  
    Despite the proposed approach generally outperforms the ALOHA-like approach of LoRaWAN in term of transmission success probability, from Fig.~\ref{fig:ef} we can conclude that for applications requiring sporadic transmission of small payloads, \mbox{S-LoRa} can lead to a reduction in energy efficiency, even in the case of very crowded LoRa networks. This is especially the case for the devices that, thanks to an advantageous link budget, can transmit at higher data rates using the smaller SFs.

	\section{Conclusion}
	\label{sec:conclusions}
	Currently, LoRaWAN represents one of the most widely adopted LPWAN technology, capable of providing the long-range wireless connectivity required by many IoT applications. The MAC layer in LoRaWAN, like that of many other LPWANs, is based on pure ALOHA for minimizing the traffic overhead and for coping with duty-cycle limitation. 
    To improve the scalabilty of a single LoRaWAN gateway, this work proposed the use of an improved LoRaWAN MAC scheme based on slotted ALOHA (\mbox{S-LoRa}). Compared to  related work in the literature, the main contribution is the use of an out-of-band synchronization based on \mbox{FM-RDS} broadcasting, a solution that natively covers wide areas both indoor and outdoor.
    The paper provides an analytical model and an experimental characterization of the synchronization performance of the proposed approach. 
    An extensive set of simulations, parameterized accordingly to a smart city scenario (e.g., smart metering), was carried out to study the proposed communication mechanism. 
    In a dense network with 10000 devices, the proposed scheme shows an improvement of up to 100\% in terms of success probability compared to the standard LoRaWAN approach. In addition, in terms of energy efficiency, although the proposed \mbox{S-LoRa} communication can have a negative gain in some conditions, it remains effective for relatively short transmission intervals and large message sizes. As a concluding remark, it has to be highlighted that the proposed time slot arrangement (actually constituting what is normally defined as a superframe), can be modified in order to include both uplink and downlink periods. Future works are planned for evaluating hybrid in-band and out-of-band sync strategies, depending on the overhead in terms of lost communication opportunities and increased power consumption for possible overhearing.
    

	
	%

	



	\ifCLASSOPTIONcaptionsoff
	\newpage
	\fi

	
	
	\bibliographystyle{IEEEtran}
	\bibliography{IEEEabrv,biblio}
\end{document}